\RequirePackage{amsmath}
\RequirePackage{wrapfig}
\documentclass[runningheads]{llncs}
\usepackage{amssymb}
\usepackage{booktabs}
\usepackage[algoruled,linesnumbered]{algorithm2e}
\usepackage{color}
\usepackage{cite}
\usepackage{enumitem}
\usepackage{floatrow}
\usepackage{graphicx}
\usepackage{hyperref}
\usepackage{mathtools}
\usepackage{MnSymbol}
\usepackage{placeins}
\usepackage{nag}
\newfloatcommand{capbtabbox}{table}[][\FBwidth]
\begin{document}
	\title{Technical Report: Refining Case Models Using Cardinality Constraints}
	\author{Stephan Haarmann$^1$ \and
	Marco Montali$^2$\and
	Mathias Weske$^1$}
	\institute{$^1$Hasso Plattner Institute, University of Potsdam, Potsdam, Germany
		\email{\{stephan.haarmann;mathias.weske\}@hpi.de}\\
		$^2$Free University of Bozen-Bolzano, Bolzano, Italy\\
		\email{montali@inf.unibz.it}}
	\authorrunning{S. Haarmann et al.}
	\titlerunning{TR: Refining Case Models Using Cardinality Constraints}
	\maketitle
		\SetAlgoLined
		\SetAlFnt{\scriptsize}
		\SetKwInOut{Input}{Input}
		\SetKwInOut{Output}{Output}
		\renewcommand{\sectionautorefname}{Section}
		\renewcommand{\figureautorefname}{Fig.}
		\renewcommand{\algorithmautorefname}{Algorithm}
	\begin{abstract}
Traditionally, business process management focuses on structured, imperative processes. With the increasing importance of knowledge work, semi-structured processes are entering center stage. Existing approaches to modeling knowledge-intensive business processes use data objects but fail to sufficiently take into account data object cardinalities. Hence, they cannot guarantee that cardinality constraints are respected, nor use such constraints to handle concurrency and multiple activity instances during execution. This paper extends an existing case management approach with data object associations and cardinality constraints. The results facilitate a refined data access semantics, lower and upper bounds for process activities, and synchronized processing of multiple data objects. The  execution semantics is formally specified using colored Petri nets. The effectiveness of the approach is shown by a compiler translating case models to colored Petri nets and by a dedicated process execution engine.
\end{abstract}
	\section{Introduction}
\label{sec:introduction}

Organizations apply Business Process Management (BPM) to specify, organize, analyze, and enact business processes.
Models play an important role in documenting, improving, configuring, and monitoring these processes.
Control flow-oriented process modeling languages, such as BPMN~\cite{BPMN}, are suited for well-structured processes. However, they lack support for semi-structured ones often required for knowledge work~\cite{diCiccio2015}.
When executing knowledge-intensive processes, knowledge workers make informed decisions choosing from a set of possible continuations for a process.
Thereby, they consult and maintain data objects.

Effective models of knowledge-intensive processes must capture flexible and data-centric behavior concisely and comprehensibly. Still, knowledge workers must decide how to progress each case in an ad-hoc manner.
To address these requirements,  among others, declarative~\cite{pesci2007,hildebrandt2010,aalst2017} and artifact-centric modeling approaches~\cite{Hull08} have been proposed.
Declarative approaches concisely define processes with many variants through rules that eliminate undesired behavior.
Artifact-centric models decompose processes based on data objects, called artifacts, or states thereof and combine the parts dynamically at run-time.

There also exist hybrid approaches such as fragment-based Case Management (fCM)~\cite{meyer2014,hewelt2016}.
fCM combines traditional BPMN-like control flow with a stronger focus on data objects.
Processes are defined through multiple activity-centric process fragments. 
Each case manages a set of data objects, which can be accessed by all fragments of a given case. Consequently, the data requirements for process activities can lead to sophisticated dependencies among fragments. Data objects and their states are not only used for determining the process execution semantics but also for defining the goal of a case.

While previous works on fCM have focused on the approach from a conceptual perspective, process execution semantics have only been investigated informally.
Behavior is specified in fragments, which are loosely coupled through data requirements and shared objects.
Notably, these dependencies are not merely based on data objects and their states but also by their mutual associations and their cardinalities.
In this paper, we focus on the following research question:
\begin{quote}
	\emph{RQ:}\quad How can we formally characterize the execution of fCM models while considering data associations and cardinality constraints?
\end{quote}
We present a thorough execution semantics for fCM case models, which takes into account process fragments and data objects with their respective states.
Cardinality constraints of data objects are captured in the domain model and considered in activities' enablement conditions.
The formalization is expressed as a translational semantics mapping fCM case models to colored Petri nets (CPN).
It allows us to describe the process behavior precisely, to verify models formally, and to provide IT support during execution.
In this regard, we present prototypes of a compiler (fCM to CPN) and a dedicated execution engine.

In \autoref{sec:preliminaries}, we present the fCM language conceptually and by an example.
We specify the execution semantics in \autoref{sec:execution_semantics}.
We present our prototypes and a discussion in \autoref{sec:discussion}.
\autoref{sec:related_work} discusses related work.
We conclude our paper in \autoref{sec:conclusion}.
	\section{Concepts and Running Example}
\label{sec:preliminaries}
An fCM case model contains a domain model, an object life cycle (OLC) for each class in the domain model, process fragments, and termination conditions~\cite{hewelt2016}.
We extend case models with data associations and cardinality constraints and thus get a full-fledged but non-hierarchical data model to represent structural aspects.
In this section, we define case models formally and illustrate the concepts with an example.

There is exactly one domain model (\autoref{def:domain_model}) for each case model.
It includes classes, associations, and cardinality constraints.
For the sake of the mapping's brevity, we do not consider generalization and specialization.
We depict domain models as UML class diagrams~\cite{UML} with two sets of cardinality constraints.
\emph{Global cardinality constraints} must always hold. They are analogue to UML's multiplicities.
\emph{Goal cardinality constraints} must only hold when the case terminates---in the diagram, they are preceded with $\diamond$.
Each domain model has a single dedicated class for case objects.
This class is instantiated once and only once for each case.

\begin{figure}[htb]
		\centering
		\includegraphics{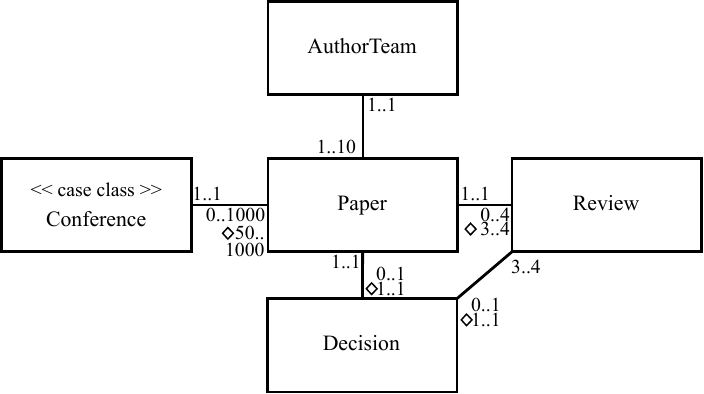}
		\caption{A class diagram describing the domain model of the submission and reviewing phase of an academic conference. Pairs of classes with positive upper bounds for the corresponding cardinality constraints are associated in the class diagram. The corresponding cardinality constraints are annotated. Goal cardinality constraints are annotated with a leading $\diamond$ if and only if they refine the corresponding global constraints.}
		\label{fig:domain_model}
\end{figure}

Consider the process of submitting and reviewing papers at an academic conference.
The domain model (\autoref{fig:domain_model}) contains classes for the conference, the papers, author teams, reviews, and decisions, as well as the depicted associations.
Cardinality constraints refine the association: a conference may have up to 1000 papers, and for each paper, there is exactly one conference (global constraints).
Likewise, the goal cardinality constraints define ranges that eventually must be satisfied: although a conference does not require any papers, it must eventually have 50 or more.
A goal cardinality constraint may only refine the lower bound of the respective global cardinality constraint.
Due to the fact that objects cannot be deleted (in fCM), the upper bound is the same for both types of constraints.
Note, we only support binary associations and only one association between a pair of classes.
Furthermore, all associations must be existential.
This means given two classes $c_1$ and $c_2$ it holds that if $u(c_1,c_2) > 0 \Rightarrow l(c_1,c_2) > 1 \vee l(c_2,c1) > 1$.
Finally, we do not support many to many associations:
\[u(c_1,c_2) > 1 \Rightarrow u(c_2,c_1) = 1\]
Incompatible domain models can be made compliant using standard reification techniques (see \autoref{sec:execution_semantics}).
\begin{definition}[Domain Model]
	\label{def:domain_model}
	A domain model is a tuple $D=(C,c_{co},l,\diamond l,u)$ where
	\begin{itemize}
		\item $C$ is a finite non-emtpy set of classes for the involved data objects;
		\item $c_{co}\in C$ is the case class describing the case object identifying a case;
		\item $l,\diamond l,$ and $u$ are functions mapping a pair of classes to an unsigned integer specifying the domain model's cardinality constraints' global lower bound, goal lower bound, and global upper bound, respectively.
	\end{itemize}
	Given two classes $c_1,c_2\in C$, $l(c_1,c_2)$ specifies the number of $c_1$ objects each $c_2$ object is associate to at least \emph{(}lower bound of the global cardinality constraint\emph{)}.
	The value $u(c_1,c_2)$ specifies the corresponding upper bound. The two classes are associated if $u(c_1,c_2)>0$ \emph{(}and $u(c_2,c_1)>0$\emph{)}.
	The function $\diamond l$ 	specifies the goal cardinality constraint.
	The value $\diamond l(c_1,c_2)$ defines the number of $c_1$ objects every $c_2$ object must be associated to eventually \emph{(}goal lower bound\emph{)}.
	For all classes $c_1,c_2\in C$ the following properties must hold:
	\begin{itemize}
		\item $u(c_1,c_2)>0 \Leftrightarrow u(c_2,c_1)>0$
		\item $l(c_1,c_2)\leq \diamond l(c_1,c_2)\leq u(c_1,c_2)$
	\end{itemize} 
\end{definition}

\begin{figure}[htb]
	\centering
	\includegraphics[]{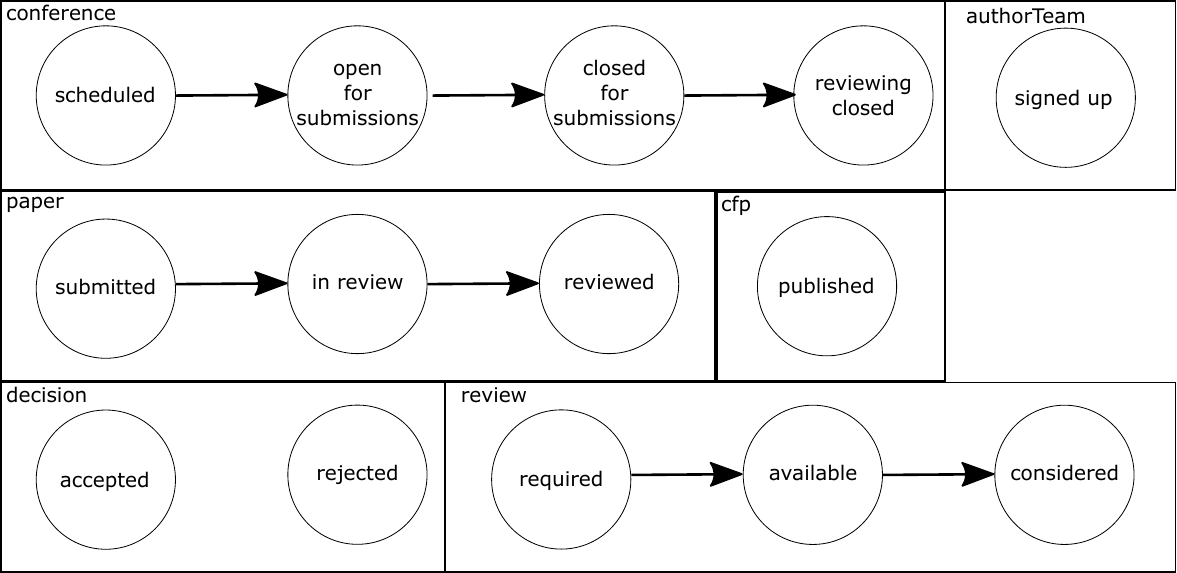}
	\caption{Object life cycles for each class (see \autoref{fig:domain_model}) in the conference example depicted as graphs. Each object life cycles consists of a set of states and state transitions between these states. States are represented by labeled circles and transitions by unlabeled arcs.}
	\label{fig:olcs}
\end{figure}

For each class in the domain model, the case model has one object life cycle (OLC) (\autoref{def:object_life_cycle}).
An OLC is a state transition system containing a finite set of states and state transitions.
The OLC defines a space in which activities can be defined: they may alter multiple data objects according to the respective OLCs.
We visualize OLCs as graphs (cf. \autoref{fig:olcs}).
The OLC for the class Conference in the example contains the states scheduled, open for submission, closed for submissions, and reviewing closed arranged in a sequence.
OLCs may also branch, in case of alternative state progressions, or contain disconnected parts, in case of alternative initial states.
We do not define initial and final states as they can be inferred from the case model's fragments.
\begin{definition}[Object Life Cycle]
	\label{def:object_life_cycle}
	Let $Q$ be a set of states and $\xrightarrow{Q} \subseteq Q^2$ a set of state transitions.
	Then $(Q,\xrightarrow{Q})$ is a state transition system.
	An object life cycle is a state transition system for objects of a specific class.
	$U_Q$ denotes the universe of all state transition systems.
\end{definition}

Data objects and their states play a fundamental role in fCM.
We call a pair of a class and a corresponding state an \emph{object configuration}.
Based on these configurations, we define \emph{data conditions} (\autoref{def:data_condition}), which are used to specify preconditions of activities and the case models' termination conditions.
\begin{definition}[Object Configuration, Data Condition]
	\label{def:data_condition}
	Let $C$ be a set of classes and $\lambda_C:C\to U_Q$ a function assigning each class an object life cycle.
	Given a class $c\in C$ we write $\lambda_C(c).Q$ to denote the set of states for $c$.
	For a state $q\in \lambda_C(c).Q$, $(c,q)$ is called an object configuration. We also write $c[q]$ do denote such a pair.
	A data condition $d\subseteq \bigcup_{c \in C} (\{c\} \times \lambda(c).Q)$ is a set of object configurations.
	We write $U_d(C,\lambda_C)$ to denote the universe of data conditions for a set of classes $C$ with the object life cycle assignment $\lambda_C$.
\end{definition}

Objects are created and updated by activities.
Activities are contained in process fragments (\autoref{def:fragment}).
Fragments are acyclic control flow graphs similar to BPMN processes.
A fragment contains a set of activities, XOR-gateways, and optionally a start event connected through control flow.
Fragments with a start event are initial: the event starts a new case.
Although loops and AND-gateways are not supported as direct constructs, concurrency, loops, and even decisions can be realized by multiple concurrent, repeatable, and alternative fragments, respectively.
This is a deliberate design decision of fCM, since knowledge-intensive processes may differ from case to case.
For each case, knowledge workers decide which activities need to be executed, in which order, and how often.
Traditional notations, such as BPMN, often produce labyrinthine models when faced with so many decisions.
In fCM, activities read and write data objects in specific configurations partitioned into input and output sets, respectively.
Furthermore, activities can read sets of objects of the same class and in the same state.
\begin{definition}[Fragment]
	\label{def:fragment}
	Given a set of classes $C$ and a function $\lambda_C$ assigning object life cycles to the classes.
	A fragment is a tuple $f=(N,\xrightarrow{CF},i,o)$ where
	\begin{itemize}
		\item $N=A\cupdot G \cupdot S$ is the set of control flow nodes partitioned into a non-empty finite set of activities $A$, a finite set of exclusive gateways $G$, and a finite set of start events $S$. The sets $A$, $G$, and $S$ are pairwise disjoint.
		
		\item $\xrightarrow{CF}\subset (S \times (G \cup A)) \cup (A \cup G)^2$ is the control flow relation. For a control flow node $n\in N$, we write $\bullet n$ to denote the set of predecessors $\bullet n=\{n'\in N | (n',n) \in \xrightarrow{CF}\}$ and $n\bullet$ to denote the set of successors.
		
		\item $i: A \to 2^{IO}$ assigns each activity a set of input sets. The set $\bigcup_{c \in C} (\{c\} \times \lambda_C(c).Q \times \{true,f\!alse\})$ contains object configurations with additional set identifiers. If the boolean is true, the tuple refers to a list of objects in the specified configuration otherwise to a single object. A single input- or output set $s$ is a subset $s\subseteq \bigcup_{c \in C} (\{c\} \times \lambda_C(c).Q \times \{true,false\})$. $IO$ is the set of such subsets (power set).
		
		\item $o: (S \cup A) \to 2^{IO}$  assigns each activity and start event a set of output sets.
	\end{itemize}
	Furthermore, every fragment must have the following properties:
	\begin{itemize}
		\item $\xrightarrow{CF}$ is acyclic\footnote{Loops can be realized through repeatable fragments}
		\item $\forall a\in A:|\bullet a|\leq 1$ activities have at most one incoming control flow;
		\item $\forall n \in (A\cup S): |n\bullet|\leq 1$ start events and activities have at most one outgoing control flow;
		\item $\forall g \in G : |\bullet g|>0$ fragments do not start with gateways;
		\item $|S| \leq 1$ a fragment has at most one start event;
		\item $S = \emptyset\Rightarrow \exists!a\in A: \bullet a = 0$ if there is no start event, the fragment starts with a single activity.
	\end{itemize}
\end{definition}

\begin{figure}[p]
	\centering
	\includegraphics[width=\textwidth]{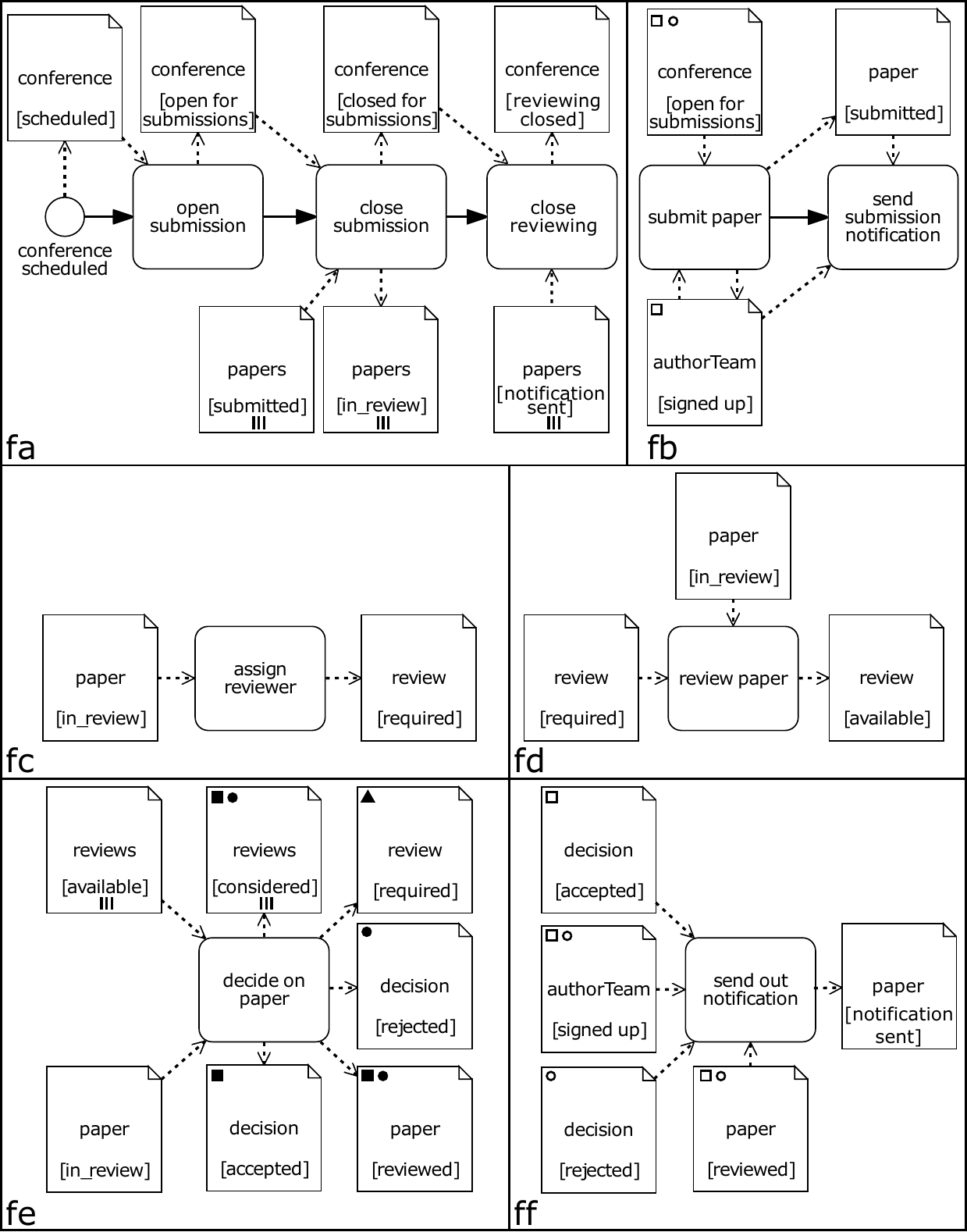}
	\caption{Process fragments for the conference example. Each fragment is depicted as graph consisting of BPMN elements. If an activity $a$ has more than one input- or output set ($|i(a)|>1$ or $|o(a)|>1$), we added markers to the object nodes to indicate the sets (this is represented by attributes in BPMN~\cite{BPMN}), e.g., submit paper has the input sets $\square=$\textit{\{(conference,open for submissions,false)\}} and $\circ=$\textit{\{(conference,open for submissions,false),(authorTeam,signed\_up,false)\}} differentiating between an author team's first and any consecutive submission. This is visually indicated by the markers $\square$ and $\circ$, respectively.}
	\label{fig:fragments}
\end{figure}

The fragments for the example are depicted in \autoref{fig:fragments}.
\emph{fa} is the initial fragment reflecting important milestones of a case, \emph{fb} is for submissions, \emph{fc} for assigning reviewers, \emph{fd} for creating reviews, \emph{fe} for deciding on a paper, and \emph{ff} for sending notifications. Fragment \emph{fb--ff} can be repeated multiple times. Instances of fragment \emph{fc--ff} can run concurrently and in various orders. 

A case can terminate if one of the termination conditions and the goal cardinality constraints are satisfied.
A termination condition is a data condition (see \autoref{def:data_condition}).
Each case model has a non-empty set of termination conditions.
The example has a single termination condition requiring the conference object in state reviewing closed:
\[\{\{(Conference, reviewing~closed)\}\}\]

A case model (\autoref{def:casemodel}) consists of a domain model, a set of fragments, an OLC for each class in the domain model, and a set of termination conditions.
The domain model is a central element defining the classes of data objects and additional constraints for all other parts.
\begin{definition}[Case Model]
	\label{def:casemodel}
	A case model is a tuple $M=(D,\lambda_C,F,T)$ where
	\begin{itemize}
		\item $D=(C,c_{co},l,\diamond l,u)$ is a domain model;
		\item $\lambda_C:C \to U_Q$ assigns each class of $D$ an object life cycle;
		\item $F$ is a set of fragments defined over $C$ and $\lambda_C$; and
		\item $T\subseteq U_d(C,\lambda_C)$ is a non-empty set of data conditions called termination conditions that specify the goal of every case.
	\end{itemize}
\end{definition}

We, furthermore, define criteria for well-formed models that assert that activities can be executed.
An object of $c_2$ depends on another object of $c_1$ if $l(c_1,c_2)>0$.
We call the object of class $c_2$ a \emph{dependent} and the object of $c_1$ a \emph{supporter}.
The \emph{dependent} object must be created within the context of the \emph{supporter}.
This means, that the \emph{supporter} object must be read or co-created when the \emph{dependent} object is created.
It can be that two objects mutually depend on one another (1-to-1 associations).
In this case, they must be co-created.
Activities that would create a \emph{dependent} without its \emph{supporter} would violate global cardinality constraints and could not be executed.

Consider the classes Conference and Paper.
Paper is the \emph{dependent} and Conference the \emph{supporter}.
In case of AuthorTeam and Paper, AuthorTeam can be both \emph{supporter} and \emph{dependent}.
This is reflected in the two input sets of activity ``submit paper`'': if a new AuthorTeam object is created, the paper is co-created (both are dependent/supporter of the other).
However, future submissions of the same team can be created without instantiating the AuthorTeam again.

If two classes $c_s$ and $c_r$ have a lower bound greater than one, $l(c_s,c_r)>1$ then $c_r$ requires a set of $c_s$ objects.
This must be respected by the fragments: an activity that instantiates $c_r$ must read a set of $c_s$ objects.
While sets of objects can be read, objects of a specific type can only be created one at a time.
In the example, a decision depends on three to four reviews.
Thus ``decide on paper'' must read a list of review objects.

Whenever a set of data objects is read, there must be a single reference object.
A reference object is an object (not a set thereof) associated with all the objects consumed in a set.
We use the associations to selectively operate on those objects associated with the reference object.
In case of activity ``decide on paper'' the paper is the reference object.
Sets can also be used for batch processing apart from lower bounds greater than one.
In the example, the activity ``close submission'' reads all papers for a given conference and updates them respectively.
	\section{Execution Semantics}
\label{sec:execution_semantics}

Formal execution semantics are the foundation for execution engines and process verification, monitoring, and analysis.
In this section, we provide a formalization of fCM case models by means of a translational semantics to colored Petri nets (CPNs).
However, first, we walk through an execution of the example process.
Next, we provide the formal semantics.

\subsection{Walk-Through}

A new case of the example in \autoref{sec:preliminaries} begins once a conference is scheduled (\emph{fa}).
The start event creates the case object (conference).
In general, non-initial fragments can begin after the start event; in the example, however, no fragment can be instantiated yet because the data prerequisites of their respective first activities are not satisfied.
Thus, the conference must be opened for submissions first (\emph{fa}) to fulfill the requirements of ``submit paper'' (\emph{fb}).

When an author team submits their first paper (input set $\bullet$), a paper object and an author team object are created.
Both get associated since objects that are created together and whose classes are associated get associated.
Papers also get associated with the conference data object since objects read get associated with newly created objects if their classes are associated.
Global cardinality constraints must still hold, i.e., an author team may submit multiple papers but not more than ten.
Papers can be submitted concurrently (multiple instances of \emph{fb}), but an instance of \emph{fb} handles the submission and notification of only one paper.

The submission is closed eventually (\emph{fa}).
Since no new papers can be accepted anymore, the conference-to-paper associations must meet the goal cardinality constraint.
So, activity ``close submission'' requires 50 or more papers and updates all papers for the conference.

After the submission has been closed, reviews for each paper are assigned (\emph{fc}) and subsequently created (\emph{fd}).
At most four reviews can be assigned to each paper (global cardinality constraint).
Once all reviews for a given paper have been created, fragment \emph{fe} can be executed:
if there are less than three reviews, the outcome is an additional required review (output set $\blacktriangle$);
if there are three reviews, the knowledge workers may decide that an additional review is needed or accept/reject the paper (output sets $\bullet$ and $\blacksquare$, respectively);
if there are already four reviews, the first option ($\blacktriangle$) is no longer applicable.
Once a decision for the paper has been created, a notification is sent to the authors (\emph{ff}).
Once notifications for all papers have been sent, the reviewing phase is closed (\emph{fa}).
The whole case can be closed when the termination condition \emph{conference}[\emph{reviewing closed}] is fulfilled and all goal cardinality constraints are satisfied.

\subsection{Formalization}
The behavior of a case model is defined jointly by all its parts:
the fragments layout the general behavior, which is refined with constraints from the OLCs and the domain model;
eventually, a case meets a termination condition and can be closed.
We map fCM models to CPNs to describe the execution semantics formally.

\paragraph{Assumptions \& Preprocessing.}
We require that domain models satisfy certain constraints.
Every association is existential (A1).
An association is existential when \textit{at least one} of the corresponding global cardinality constraints has a positive lower bound.
So, an object of one class cannot be created without an object of the other.
We call the objects \emph{dependent} and \emph{supporter}, respectively.
Furthermore, many-to-many associations are not supported (A2).
A domain model $D=(C,c_{co},l,\diamond l, u)$ meets the assumptions if the following properties are satisfied:
\begin{description}
	\item [A1] $\forall c_1,c_2 \in C: u(c_1,c_2) > 0 \Rightarrow l(c_1,c_2) > 0 \vee l(c_2,c_1 > 0)$
	\item [A2] $\forall c_1,c_2 \in C: u(c_1,c_2) > 1 \Rightarrow u(c_2,c_1) = 1$
\end{description}
By definition (cf. \autoref{def:domain_model}), at most one association exists between two classes (A3), and only binary associations are supported (A4).

\begin{figure}[htb]
	\centering
	\includegraphics[]{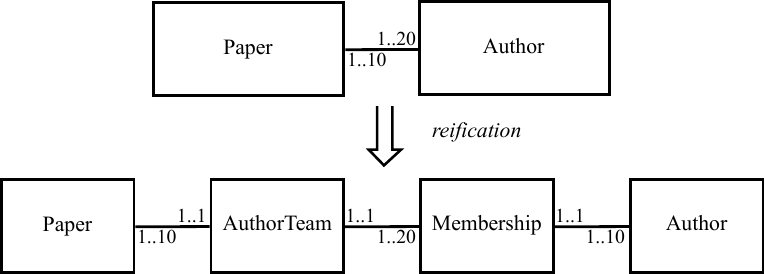}
	\caption{A paper is written by multiple authors, which may write multiple papers. This many-to-many association can be reified: we introduce a class AuthorTeam and a class Membership. An author has a membership for an author team.}
	\label{fig:reification}
\end{figure}

Domain models that break these assumptions can be transformed into compliant ones through reification.
An association violating the constraints is replaced by multiple classes and associations (see \autoref{fig:reification}).
For example, a many to many association $A$ linking the classes $c_1$ and $c_2$ can be reified into a class $c_A$ associated with both $c_1$ and $c_2$.
This comes at the price of enriching the vocabulary of the domain; thus, reification is best performed by domain experts, who provide meaningful names, cardinality constraints, and states for the added classes and who adapt the fragments accordingly.

\begin{figure}[htb]
	\centering
	\includegraphics[]{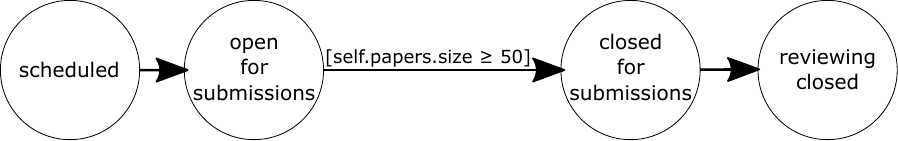}
	\caption{OLC of the conference augmented with a guard. The conference can only change to the state ``closed for submissions'' if there are at least 50 associated papers. This requirement arises from the goal cardinality constraints and the fragments: eventually, there must be 50 papers, and the fragments forbid adding papers for a conference once it is in the state ``closed for submissions''.}
	\label{fig:conditional_transition}
\end{figure}

Besides these assumptions, we perform a preprocessing step before translating the case model.
During preprocessing, we augment state transitions with guards asserting goal cardinalities.
As described, associations may be finalized before the case terminates.
In the moment of finalization, goal cardinality constraints must hold, e.g., a conference must have at least 50 associated papers when it is closed for submissions.
Associations are established when one object (dependent) is created in the context of another (supporter).
This is done by activities which make assumptions about the supporter's state, i.e., ``submit paper'' must read the conference in the state ``open for submission''.
If an activity changes the state irreversibly the goal cardinality must hold.
For each state of a supporter, we check whether the fragments allow adding associations (for a specific dependent).
If so, we check in the OLC for the outgoing state transitions whether the target state or any state reachable supports adding further associations.
If not, we add a guard to the state transitions that requires the object (of the supporter) to satisfy the goal cardinality constraint (see \autoref{fig:conditional_transition}).

\paragraph{Translation.}
A CPN is a Petri net with typed places and tokens.
The enablement of transitions may depend on the value of tokens, and the firing may derive new tokens.
As such, CPNs can represent data-based systems, i.e., fCM case models.

Types in CPNs are called colorsets.
We represent an object by an identifier (\texttt{ID}), which consists of a class and an integer (i.e., the number of existing  objects of the same class at the point of creation).
An association is an unordered pair of IDs.
We further have types for sets of IDs (\texttt{Set<ID>}) and sets of associations (\texttt{Set<Association>}).
A colorset \texttt{ControlFlow} for the control flow maps each class to an \texttt{ID} or \texttt{NULL}.
We further use places of colorset \texttt{INT} (integer numbers) and \texttt{Unit} (blank tokens).
Formally coloresets are defined as:
{
	\scriptsize
\begin{align*}
	ID &= Classes \times \mathbb{N}\\
	Association &= {{o_1,o_2} \subset ID | o_1 \neq o_2}\\
	Set<ID> &= P(ID)\\
	Set<Association> &= P(Association)\\
	ControlFlow &= Classes \rightarrow ID
\end{align*}
}

\autoref{alg:overview} outlines the mapping.
We created places for abstract states of the case (l.~1), class-specific counters (l.~3), and object configurations (l.~5).
Transitions are added for all control flow nodes (l.~12--29) and each termination condition (l.~31).
We present each of these steps, both in text and visually, in the remainder of this section.
\begin{algorithm}[ht]
	\caption{Algorithm for translating a case model to a CPN. Details are given in the figures (see comments).}
	\label{alg:overview}
	\KwData{$((C,c_{co},l,\diamond l,u),\lambda_C,F,T)$}
	createGlobalPlaces()\tcc*{\autoref{fig:global_places}}
	\ForEach{c $\in$ C}{
		createCounterPlace(c)\tcc*{\autoref{fig:data_places}}
		\ForEach{q $\in$ $\lambda_C(c).Q$}{
			createObjectConfigurationPlace(c,q)\tcc*{\autoref{fig:data_places}}
		}
	}
	\ForEach{$(N,\xrightarrow{CF},i,o) \in F$}{
		\ForEach{$(n,n') \in \xrightarrow{CF}$}{
			createControlFlowPlace($(n,n')$)\tcc*{\autoref{fig:controlflow}}
		}
		\ForEach{$n\in N$}{
			\uIf{$n$ is gateway}{
				\ForEach{$(n',n),(n,n'')\in \xrightarrow{CF}$}{
					createGatewayTransition($n,n',n''$)\tcc*{\autoref{fig:gateways}}
				}
			}\uElseIf{$n$ is start event}{
				\ForEach{$\blacksquare \in o(n)$}{
					createStartEventTransition($n,\blacksquare$)\tcc*{\autoref{fig:start_event}}
				}
			}\Else{
				\ForEach{$\blacksquare \in o(n)$}{
					\ForEach{$\square \in i(n)$}{
						createActivityTransition($n,\square,\blacksquare$)\tcc*{\autoref{fig:activity_data_flow},\ref{fig:batchprocessing}}
					}	
				}
			}
		}
	}
	\ForEach{$t \in T$}{
		createTerminationConditionTransition(t)\tcc*{\autoref{fig:termination_condition}}
	}
\end{algorithm}

\begin{figure}[htb]
	\centering
	\includegraphics[]{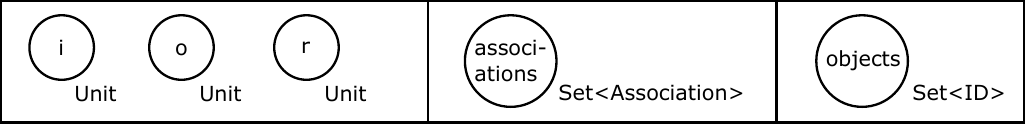}
	\caption{Global places that are created for each case model: i) an initial place \emph{i}, which holds a single token until a case is started; a place \emph{r}, which holds a single token when the case is running; and a final place \emph{o} that holds a token after the case has been closed. Furthermore, we ii) add a place \textit{associations} of type \texttt{Set$<$Association$>$} and a place \emph{objects} of type \texttt{Set$<$ID$>$}.}
	\label{fig:global_places}
\end{figure}
\begin{figure}[htb]
	\centering
	\includegraphics[]{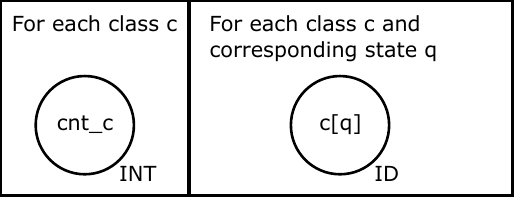}
	\caption{To support data two sets of places are added: the first set contains a place with colorset \texttt{INT} for each class; the second set contains a place of colorset \texttt{ID} for each object configuration.}
	\label{fig:data_places}
\end{figure}
\begin{wrapfigure}[8]{r}{0.4\textwidth}
	\centering
	\includegraphics[]{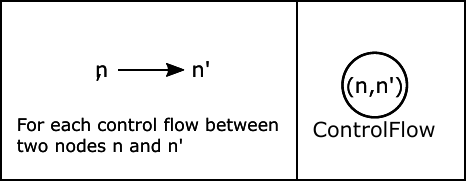}
	\caption{For each control flow arc, a ControlFlow place is created.}
	\label{fig:controlflow}
\end{wrapfigure}
The state of a CPN is defined by its current marking, the value and position of tokens in the net.
We, thus, require places for the different aspects of a case's state.
First, we distinguish between three abstract states initial, running, and terminated of the case.
We add three \texttt{Unit} places \emph{i}, \emph{r}, and \emph{o}, respectively (\autoref{fig:global_places}).
We furthermore add a single place labeled \emph{associations} with colorset \texttt{Set<Associations>} holding a single token containing all associations, and one place \emph{objects} of type \texttt{Set<Object>} holding a single token containing IDs of all data objects (\autoref{fig:global_places}).
Furthermore, a place typed \texttt{ID} for each object configuration is added (\autoref{fig:data_places}).
Additionally, we add a place of type \texttt{INT} for each class (\autoref{fig:data_places}).
These places each hold a token with initial value 0 and are labeled ``cnt\_$<$Class$>$'', e.g., \emph{cnt\_Conference}.
They are used to count the number of objects for the respective class.
Finally, a \texttt{ControlFlow} place (n,n') is added for each control flow $(n,n') \in \xrightarrow{CF}$ in each fragment (\autoref{fig:controlflow}).

Transitions are added for each control flow node and for each termination condition (\autoref{alg:overview}, l.~8--32).
A gateway is translated to one transition for each combination of incoming and outgoing control flow (l.~13--16).
A start event is translated to a transition for each of its output sets\footnote{A start event without a data output has an empty output set.} (l.~17--21).
For an activity, a transition for each combination of an input set and an output set is added (l.~22--27).
Finally, a transition for each termination condition is created (l.~30--32).

\begin{figure}[htb]
	\centering
	\includegraphics[]{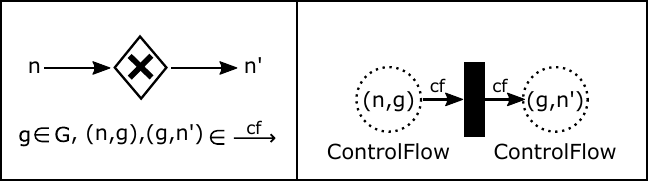}
	\caption{For each gateway we consider every combination of an incoming and an outgoing control flow. The CPN has a transition that passes the control flow for each such combination.}
	\label{fig:gateways}
\end{figure}

A transition for a gateway (\autoref{fig:gateways}) moves a token from the place representing the respective incoming control flow to the place representing the respective outgoing one.
The value is not changed, and the transition has no guards.

\begin{figure}[htb]
	\centering
	\includegraphics[]{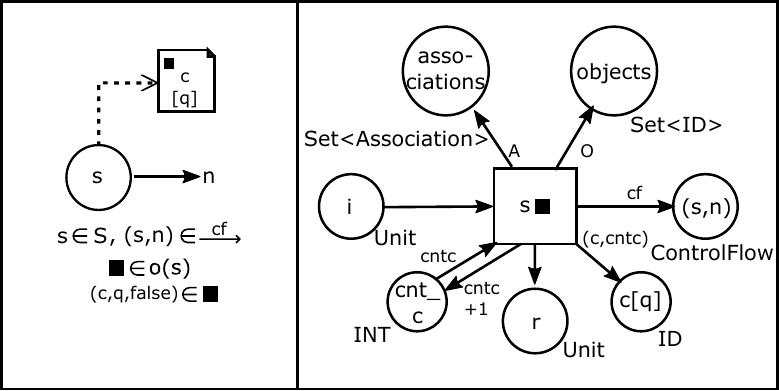}
	\caption{We create a transition for each output set ($\blacksquare$) of each start event (s). The transition moves a token from \emph{i} to \emph{r}, creates a data object for each configuration (c[q]) in the output set. In the process, the corresponding counter is incremented. The set of associations and objects is updated, and a \texttt{ControlFlow} token referencing all created objects is created. \emph{O} is a set of all objects created by the start event for the output set. \emph{A} is the set of all associations that are established.}
	\label{fig:start_event}
\end{figure}

A transition representing a start event (\autoref{fig:start_event}), takes a token from \emph{i} and produces one in \emph{r}.
If the output set represented by the transition is non-empty, counters for all objects in the output set are read and incremented.
Based on the original value of the counter new \texttt{ID} tokens for the objects are created and i) added to a set stored on the place \textit{objects} and  ii) as a single token stored on the place representing the respective object configuration.
The set of associations is initialized: a \texttt{Set$<$Association$>$} token containing associations among the created objects (if applicable) is created.
Also, if the event has an outgoing control flow, a new \texttt{ControlFlow} token is produced on the corresponding place.
The token's fields for the objects in the output set reference the created objects.

\begin{figure}[t!]
	\centering
	\includegraphics[]{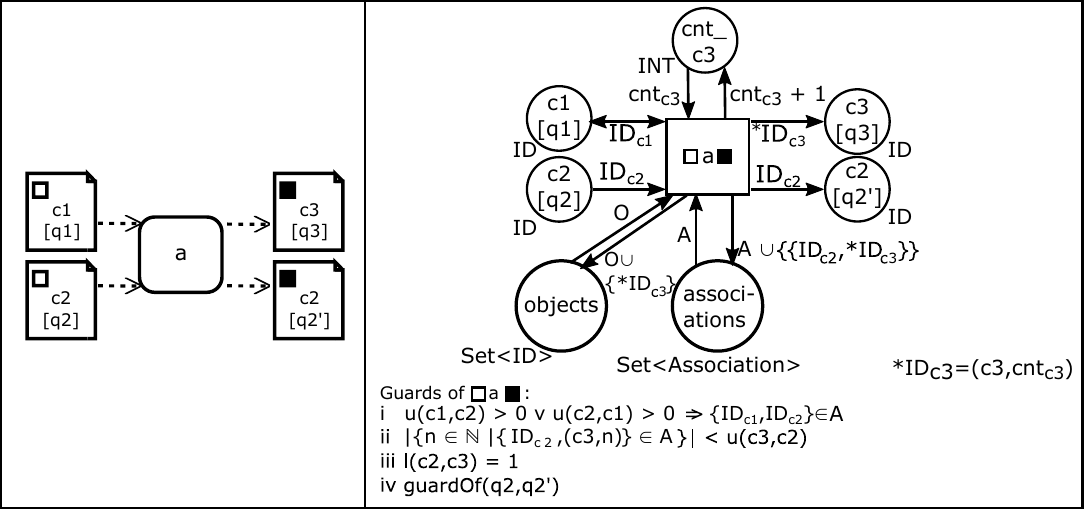}
	\caption{An activity reads objects of one input set ($\square=\{c1[q1],c2[q2]\}$) and writes objects  for one output set ($\blacksquare=\{c2[q2'],c3[q3]\}$). We assume $q2 \neq q2'$. This means objects can be read (c1,c2), updated (c2), or created (c3). The associations and references to all objects are read and updated. In this case, we assume an association between c3 and c2 but none between c1 and c3. The CPN transition consumes and produces the required tokens. The guard asserts that (i) inputs are associated if applicable, (ii) upper cardinality bounds are not exceeded, (iii) lower bounds of global cardinality constraints are satisfied, and (iv) state changes are possible. If the activity has an incoming control flow, additional guard conditions assert that data objects are consumed according to the references in the \texttt{ControlFlow} token (not depicted).}
	\label{fig:activity_data_flow}
\end{figure}

A transition for an activity (\autoref{fig:activity_data_flow}), consumes a token for each data object that is read.
It re-produces consumed tokens for each data object that is read or updated in the same or a different place, respectively.
Furthermore, new \texttt{ID} tokens are produced for newly created data objects, this includes an update of the corresponding counters, the set of objects, and the set of associations (if applicable).
New objects that may be associated (according to the domain model) to objects in the input or outputs set will be associated.
Transitions for activities without incoming control flow require a token on place \emph{r}, but the token is not removed when the transition fires.
Other activities are triggered via their incoming control flow and the respective transition require a corresponding \texttt{ControlFlow} token.
A transition representing an activity has a guard to assert
\begin{description}
	\item [i)] inputs that can be associated are associated
	\item [ii),iii)] associations that will be established by the transition do not violate cardinality constraints
	\item [iv)] the guards of the state transitions realized by the activity are satisfied
	\item [v] bindings in the control flow are not violated (not included in \autoref{fig:activity_data_flow})
\end{description}

\begin{figure}[htb]
	\centering
	\includegraphics[]{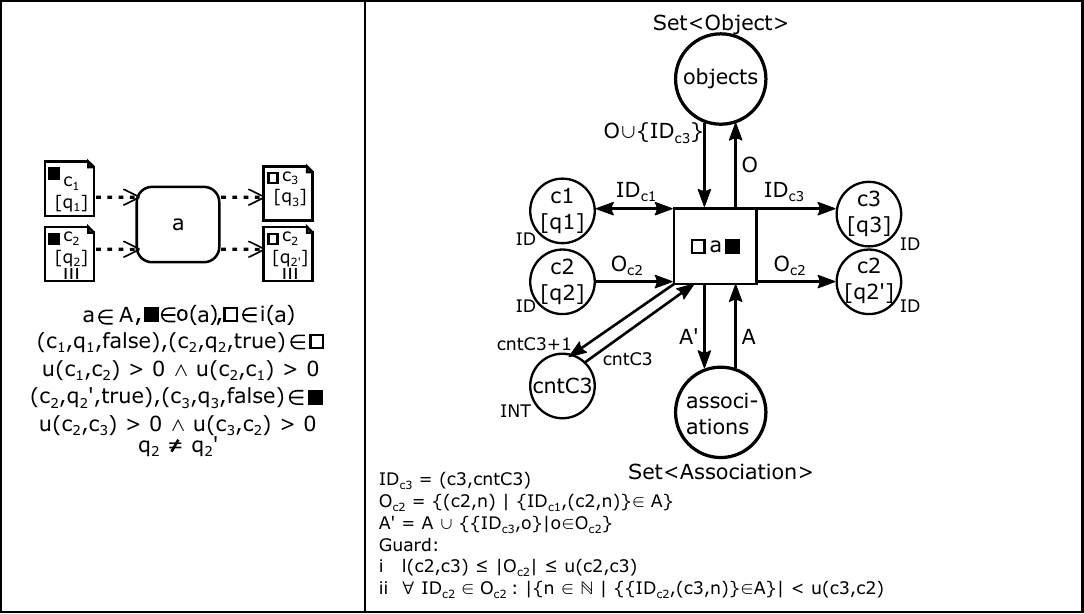}
	\caption{An activity may read and update a set of data objects. In the depicted case, a set of $c2$ objects associated with the object $ID_{c1}$ is read and updated. Furthermore, a $c3$ object is created and associated with all elements of the set. The guard asserts, that neither the global upper bounds nor the global lower bounds are violated.}
	\label{fig:batchprocessing}
\end{figure}

Activities may also represent batch behavior (\autoref{fig:batchprocessing}).
They read and possibly update multiple objects of the same type, e.g., all papers.
As described above, the set of objects is qualified by an associated reference object.
Let $\square$ be an input set containing an entry $(c_1,q_1,true)$ (a set of $c_1$ objects in state $q_1$ is read), then there exists an entry $(c_2,q_2,false)$ in $\square$ so that $u(c_1,c_2)>1$ (the two classes are associated).
If $A$ is the set of associations and $(c_2,n_2)$ the ID of the reference object, then the set $O_1$ is the set of $c_1$ objects associated with $(c_2,n_2)$:
\[O_1=\{(c_1,n_1) | \{(c_2,n_2),(c_1,n_1)\} \in A\}\]

\begin{figure}[htb]
	\centering
	\includegraphics[]{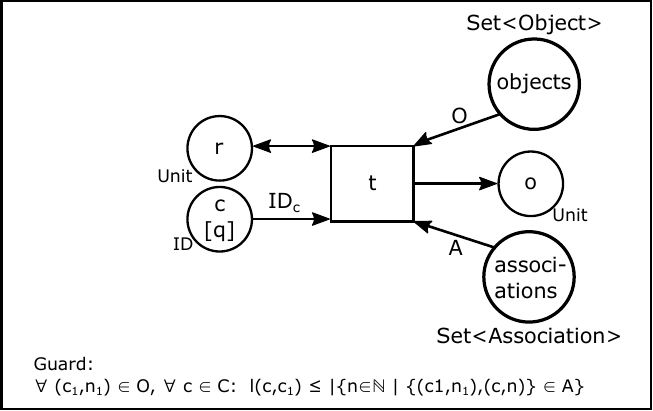}
	\caption{A termination condition $t$ contains a set of object configurations $\{c[q]\}$. The corresponding transition moves the token from \emph{r} to \emph{o},
	consumes a token for every object configuration, one for the set of objects, and one for the set of associations. The transition has a guard asserting that the goal cardinality constraints are satisfied.}
	\label{fig:termination_condition}
\end{figure}

\vspace{-2.em}
Furthermore, a transition for each termination condition is created (\autoref{fig:termination_condition}).
The transition moves a token from \textit{r} to \textit{o} terminating the case.
Furthermore, a token for each required object configuration as well as the tokens for the set of all objects and the set of associations are consumed.
Such a transition has a guard asserting that the goal cardinality constraints are met.
Let $O$ be the set of object identifiers and $A$ the set of associations.
Furthermore, let $D=(C,c_{co},l,\diamond l, u)$ be the domain model.
The formal guard condition is
\begin{equation*}
\forall (c_1,n_1) \in O, \forall c \in C:
\diamond l(c,c_1) \leq \{n \in \mathbb{N} | \{(c_1,n_1),(c,n)\}\in A\}
\end{equation*}

\FloatBarrier 
\subsection{Formalization of the Example}
Next, we formalize the example (\autoref{sec:preliminaries}).
We present the formalization one fragment at a time.
Places with the same label represent the same place, e.g., both the CPN for \emph{fa} and the one for \emph{fb} have a place labeled ``conference[open for submissions]'', but the place is created only once.

\begin{figure}[htb]
	\centering
	\includegraphics[width=\textwidth]{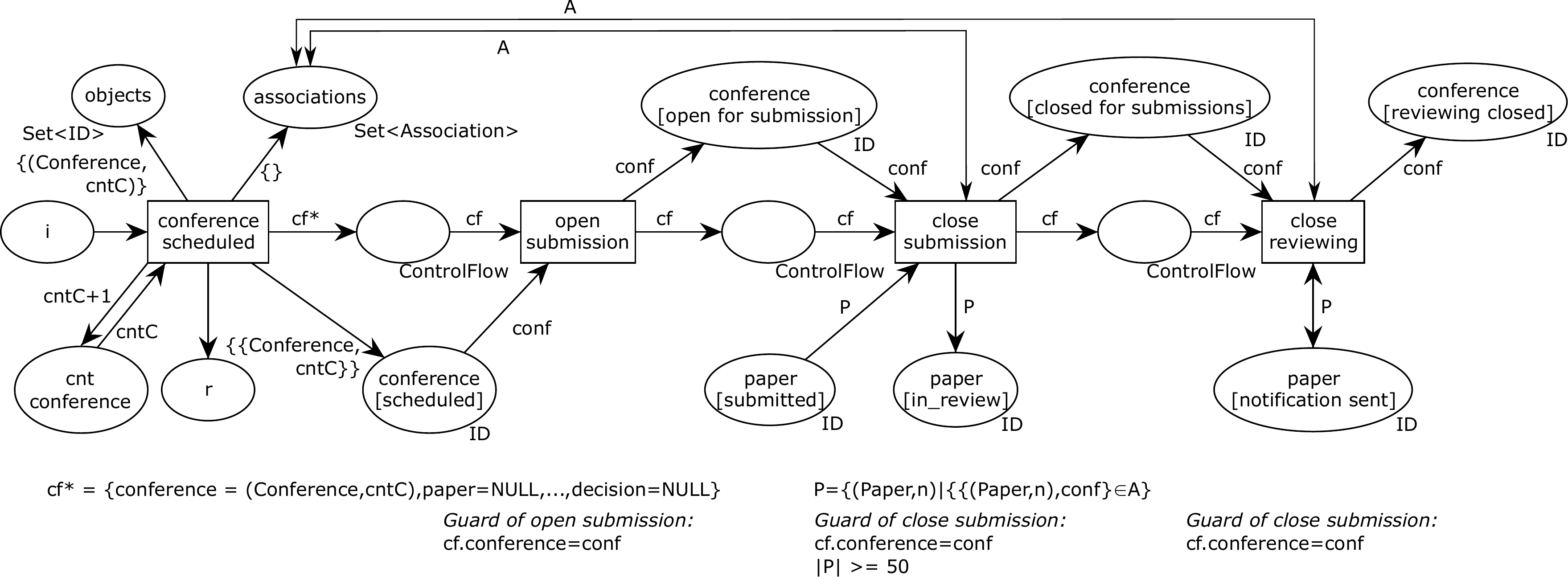}
	\caption{The formalization of fragment \emph{fa} with transitions for the start event and the subsequent activities as well as places for the general state and data objects. Places of type \texttt{Unit} have no colorset annotated. The transition ``close submission'' has a guard asserting goal cardinalities, the transitions ``close submission'' and ``close reviewing'' access a list of IDs by querying the set of associations.}
	\label{fig:fa_formal}
\end{figure}
The formalization of fragment \emph{fa} (\autoref{fig:fa_formal}) contains four transitions: one transition for the start event and one for each activity.
Neither the start event nor the activities have multiple input or output sets.
The formalization of the fragments contains places for the object configurations accessed by the control flow nodes, the counter of conference object, which is created by the start event, control flow places for all control flow arcs in \emph{fa}, and the places \emph{i} and \emph{r} denoting the abstract states \textit{initial} and \textit{running}, respectively.

The firing of the start event starts a new case.
In the CPN, a token is moved from \emph{i} to \emph{r}.
Furthermore, an \texttt{ID} token for the conference object in state \emph{scheduled} is created.
For this purpose, the counter on \emph{cnt Conference} is read and incremented.
The transition also initializes the set of associations and the set of objects (including the conference ID).
Since the start event has an outgoing control flow, the transition produces a control flow token referencing the conference. The label of \texttt{ControlFlow} places is omitted.

The control flow leads to the activity ``open submission'', which updates the state of the conference object.
The corresponding transition consumes the \texttt{ControlFlow} token and the \texttt{ID} token of the conference.
A guard asserts that the \texttt{ID} is referenced in the \texttt{ControlFlow}.
Both tokens are reproduced on the places for conference[open for submissions] and the outgoing control flow, respectively.

The next activity, ``close submission'', reads a list of papers in addition to the conference.
The corresponding transition (\autoref{fig:fa_formal}), accesses the list of associations to find all the papers associated with the conference.
All papers in the list and the conference are updated.
Therefore, the transition moves the respective \texttt{ID} tokens to the places for the corresponding object configuration.
The activity can only be executed if all papers of the conference are in state \emph{submitted} and if there are at least 50 associated papers.
This is represented in the  transition's guard.
Activity ``close reviewing'' is mapped analogously.
However, it has no successor; thus, the \texttt{ControlFlow} token is consumed but not produced.

\begin{figure}[htb]
	\centering
	\includegraphics[width=\textwidth]{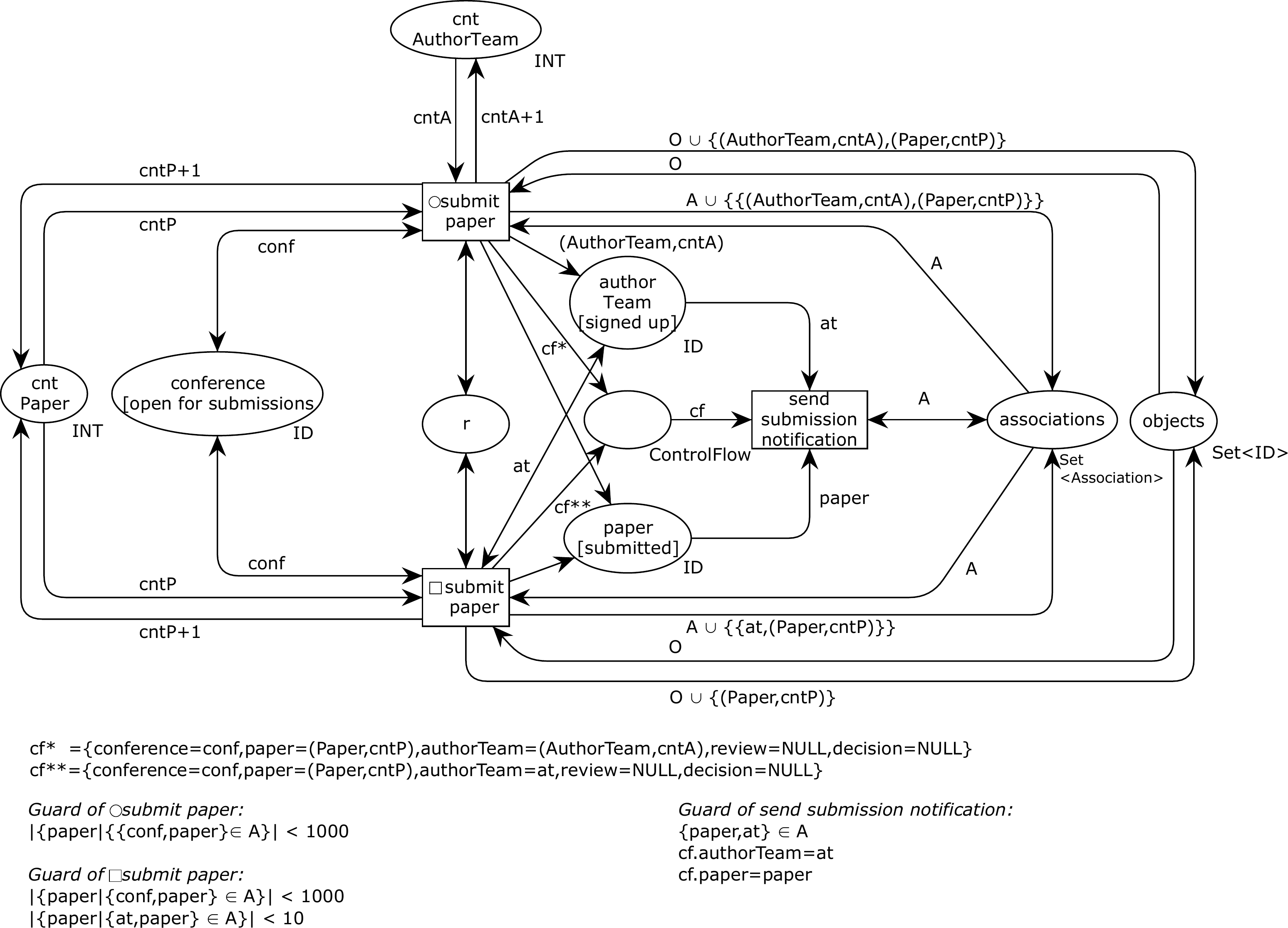}
	\caption{The formalization of fragment \emph{fb} has two transitions for activity ``submit paper'' and one for ``send submission notification''. ``Submit paper'' is mapped to two transitions because it has two input sets ($\circ$ and $\square$) and one output set. The \texttt{ControlFlow} token, that is passed from ``submit paper'' to ``send submission notification'' references the conference, author team, and paper.}
	\label{fig:fb_formal}
\end{figure}
The formalization of fragment \emph{fb} (\autoref{fig:fb_formal}) contains two alternative transitions for ``submit paper'' and a subsequent transition for ``send submission notification''.
It further contains places for the control flow, the set of associations, the set of objects, and the object configurations accessed by the activities.
The transitions for the fragment's first activity require a token on \emph{r} (the case must be running), but the token is not removed.
They, furthermore, consume and reproduce a token for a conference object in state \emph{open for submission}.
Both transitions create \texttt{ControlFlow} token that references the three involved objects: conference, author team, and paper.
Transition ``$\circ$submit paper'' represents the first submission of an author team.
An \texttt{ID} token for the author team and one for the paper is created, the IDs are added to the list of objects (O), and the pair of IDs is added to the set of associations (A).
Furthermore, an association between the conference and paper is established.
Transition ``$\square$submit paper'' is similar, but an ID of an author team is consumed and reproduced instead of created.
Both transitions must not violate cardinality constraint (see guards).
Since neither transition invalidates its precondition, \emph{fb} can be executed multiple times.
The number of executions is only limited by the upper bound of the corresponding global cardinality constraint: if 1000 papers have been submitted, the ``submit paper'' transitions are disabled.
 
The transition ``send submission notification'' consumes the previously created \texttt{ControlFlow} token.
It also consumes \texttt{ID} tokens for the referenced author team and paper (see guard).
The guard asserts that both objects are associated.
However, neither of the objects is updated; thus, the \texttt{ID} tokens are reproduced on the place they have been consumed from.
Also, neither the list of objects nor the list of associations is changed.
However, the former is accessed to check that the two objects are associated.

\begin{figure}[htb]
	\centering
	\includegraphics[scale=.5]{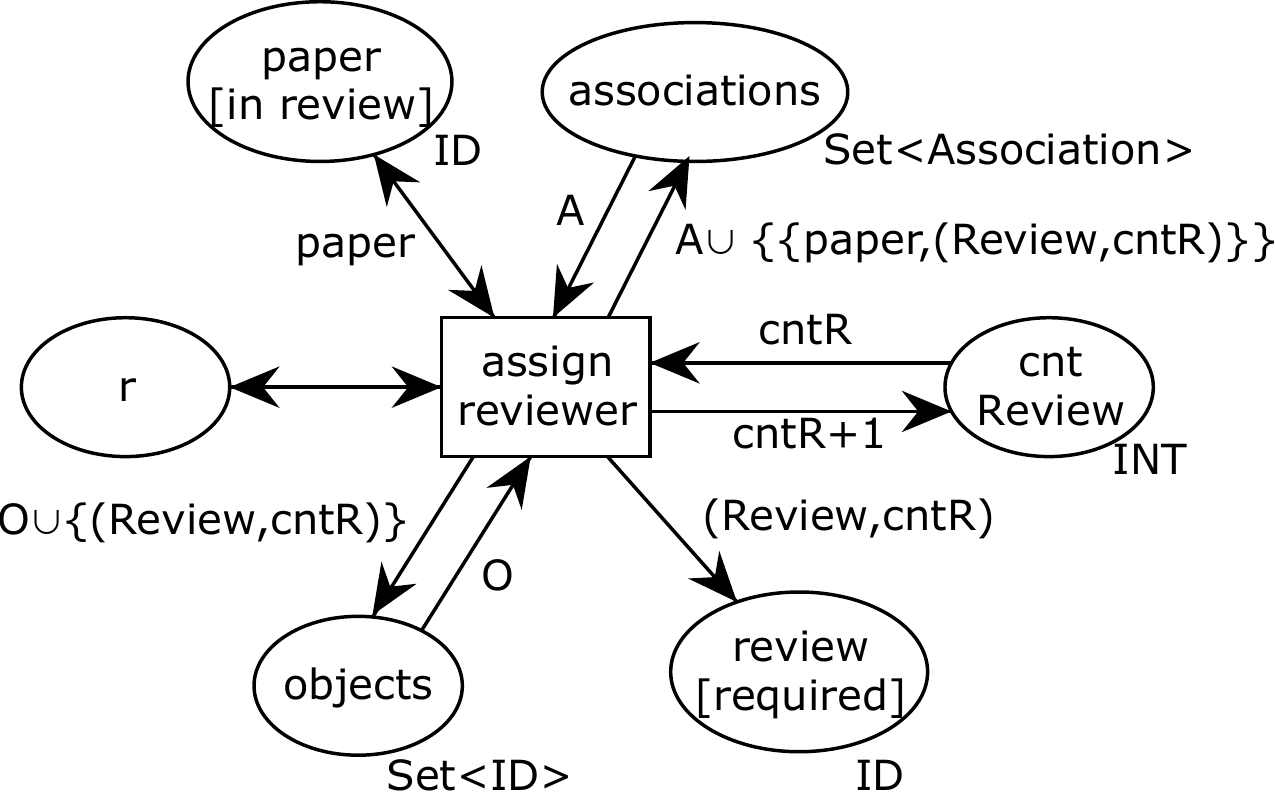}
	\caption{The CPN for fragment \emph{fc} has a single transition ``assign reviewer''. It can be executed up to four times for every submission and assigns a review to the paper.}
	\label{fig:fc_formal}
\end{figure}
Fragment \emph{fc} has a single activity with one input set and one output set.
It is mapped to a single transition (\autoref{fig:fc_formal}).
The transition requires that the case is running (token on \emph{r}).
It consumes and produces an ID token from the place ``paper[in\_review]'' and creates a new ID token on ``review[required]'' (incrementing the corresponding counter on \emph{cnt Review}).
The guard asserts that the global cardinality constraints are not violated by preventing that more than four reviews are created for each paper.

\begin{figure}[htb]
	\centering
	\includegraphics[scale=.5]{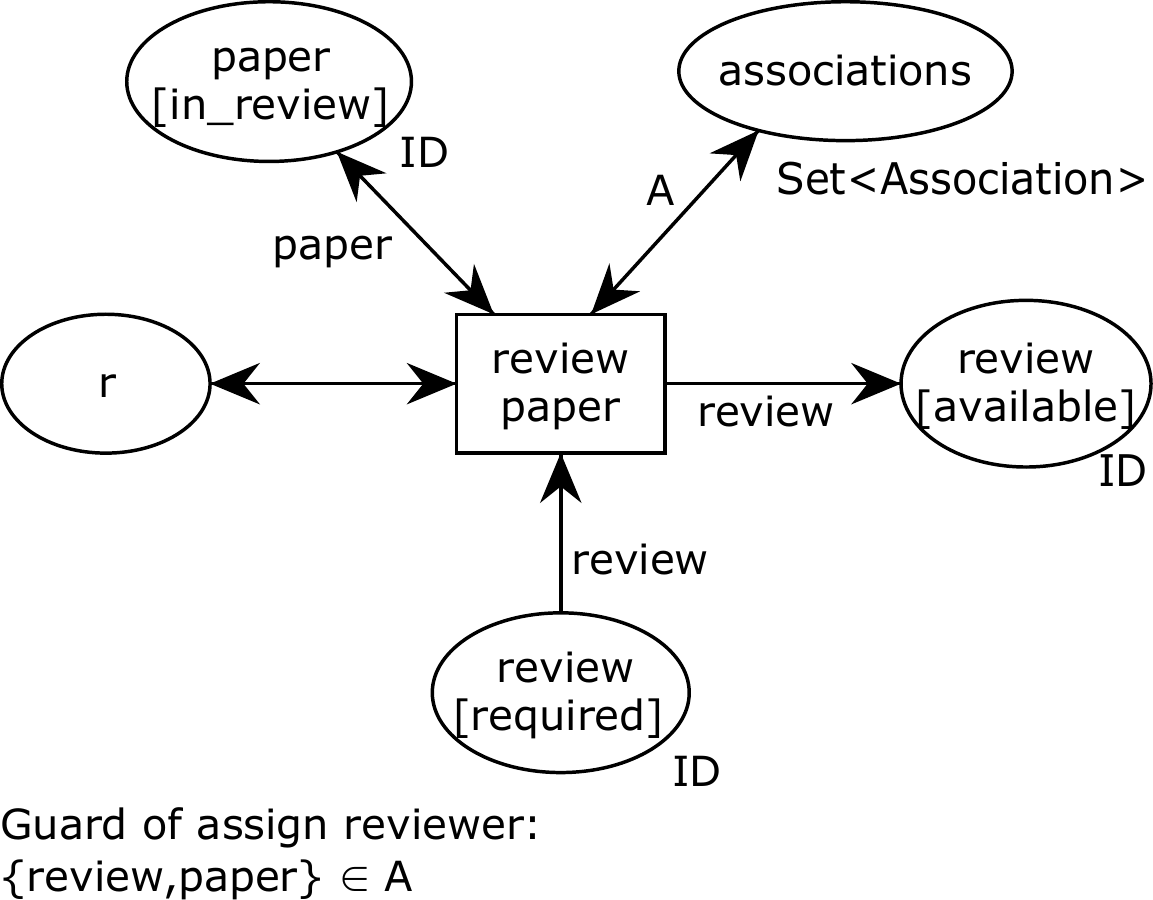}
	\caption{Formalization of fragment \emph{fd}. Reviews that are required must be created. The transition ``create review'' updates the the state of a review by moving the corresponding \texttt{ID} token from place ``review[required]'' to place ``review[available]''.}
	\label{fig:fd_formal}
\end{figure}
Fragment \emph{fd} updates a review object (\autoref{fig:fd_formal}).
Its state changes from \emph{required} to \emph{available}.
The transition ``review paper'' consumes \texttt{ID} tokens for the review and the associated paper (see guard).
While the latter remains unchanged, the \texttt{ID} token for the review is produced on the place ``review[available]''
The transition also consumes and produces a token on \emph{r} (the case must be running).

\begin{figure}[htb]
	\centering
	\includegraphics[width=\textwidth]{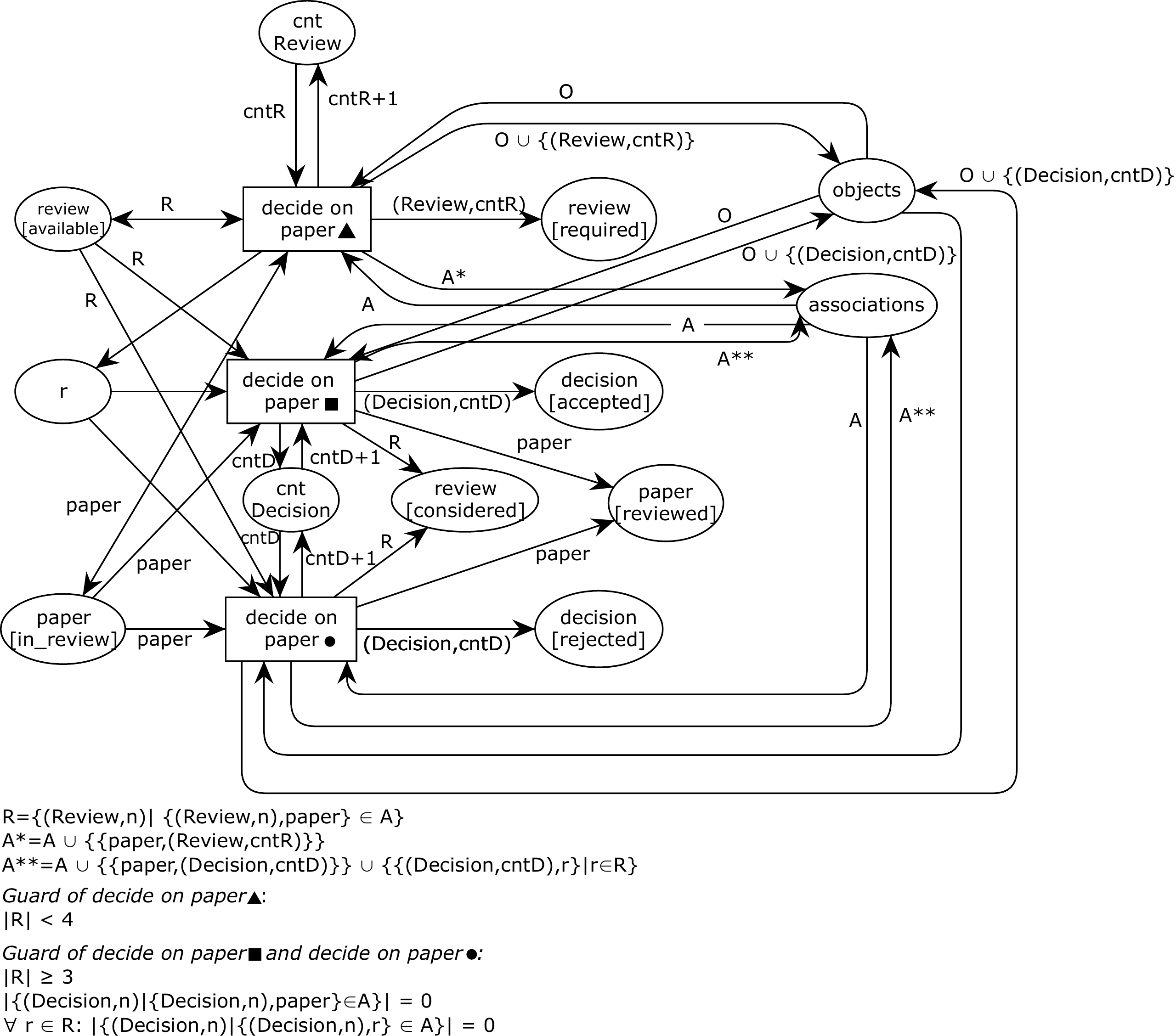}
	\caption{Formalization of fragment \emph{fe}: while the fragment has a single activity ``decide on paper'', the CPN has three corresponding transitions---one for each output set. Transition ``decide on paper$\blacktriangle$'' requires an additional review, ``decide on paper$\blacksquare$'' accepts the paper, and ``decide on paper$\bullet$'' rejects the paper. In case of acceptance or rejection, a decision object is created.}
	\label{fig:fe_formal}
\end{figure}

Fragment \emph{fe} contains a single activity with three output sets.
It is mapped to three repsective transitions (\autoref{fig:fe_formal}).
Each transition represents one of the output sets of activity ``decide on paper''.
They each consume an \texttt{ID} token for a paper in the state \emph{in\_review} and a set of IDs for the associated reviews in the state \emph{available}.
Transition ``decide on paper$\blacktriangle$'' represents the decision outcome of an additional required review.
The transition reproduces the set of \texttt{IDs} on place ``review[available]'' and the paper ID on place ``paper[in\_review]''.
It also creates a novel ID for a new review object on place ``review[required]''.
Since a new review for the paper is created, the corresponding global cardinality constraint is checked (see guard).
Furthermore, the set of object IDs and the set of associations is updated, respectively.
In case three or four reviews exist for a particular paper, it can be rejected or accepted.
These outcomes are represented by transitions ``decide on paper$\bullet$'' and ``decide on paper$\blacksquare$'', respectively.
Both transitions move the set of \texttt{ID} tokens for the reviews to the place ``review[considered]'' (updating the state of each object in the set).
Likewise, the paper's \texttt{ID} token is moved to ``paper[reviewed]''.
A new ID is created for the decision and the corresponding counter (token on ``cnt Decision'') is consumed and incremented.
The ID is added to the set of objects (O) and the set of associations (A) is updated:
the decision is associated with the paper and to each review that was considered.

Fragment \emph{fe} is a knowledge-intensive task.
The involved knowledge workers have to decide which variant fits the particular case best: e.g., which paper can be rejected or accepted, and which paper requires an additional review.
However, the cardinality constraints limit the options for the knowledge workers: papers cannot be rejected/accepted if not enough reviews exist.
On the contrary, only four reviews for a paper may exist, limiting the execution of transition ``decide on paper$\blacktriangle$''.
A semantics that does not consider cardinality constraint would allow infinite reviews and not support knowledge workers during the decision.

\begin{figure}[htb]
	\centering
	\includegraphics[scale=.5]{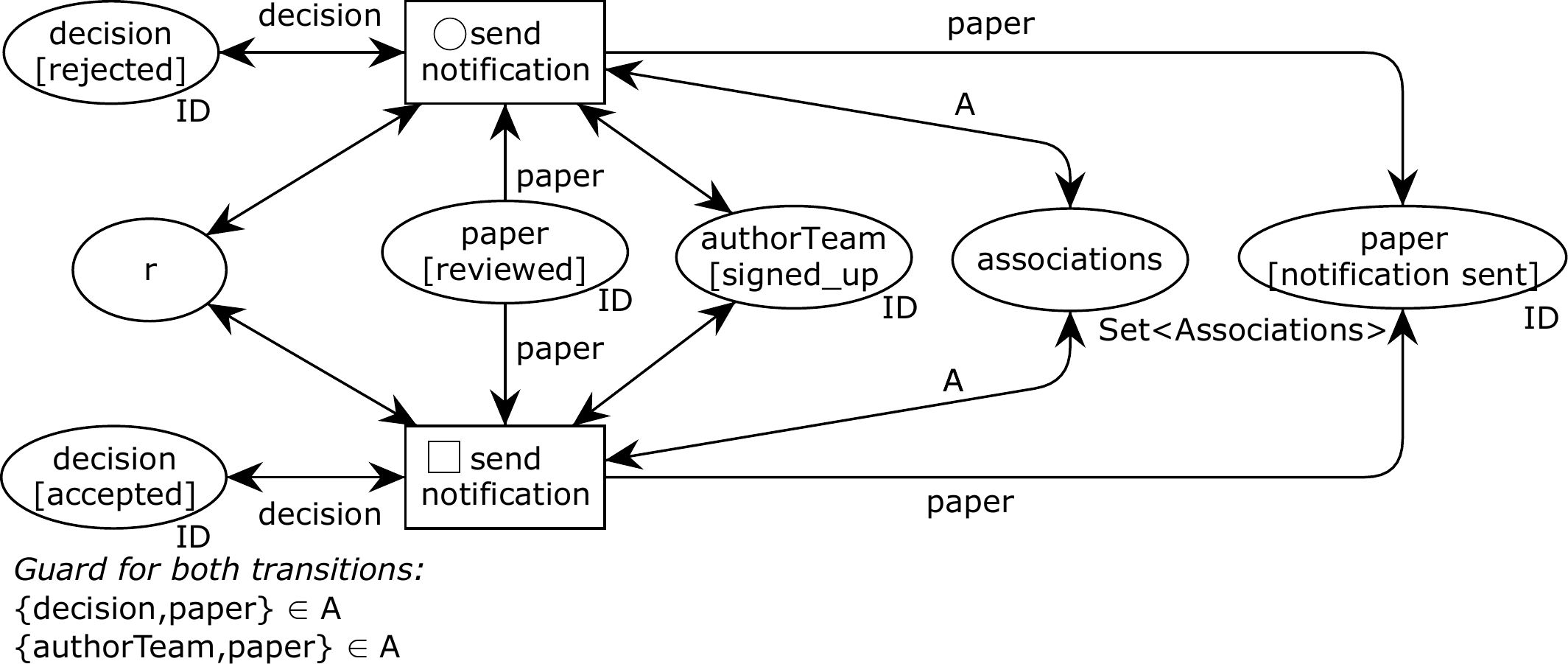}
	\caption{Formalization of fragment \emph{ff}. The two transition represent the sending of an acceptance and rejection notification ($\square$ and $\circ$). Objects are updated but not created. The set of associations is consulted to assert that the decision is associated with the paper and that the paper is associated with the author team (guard).}
	\label{fig:ff_formal}
\end{figure}
Fragment \emph{ff} contains an activity for sending notifications to the authors of a paper.
The activity updates a paper object and reads an author team object.
Furthermore, it reads the respective decision which is either in state \textit{accepted} or \textit{rejected}.
This is represented by two corresponding transition (\autoref{fig:ff_formal}).
They consume and produce \texttt{ID} tokens for the objects read, and their guards assert that the activity operates only on associated objects.

\begin{figure}[htb]
	\centering
	\includegraphics[scale=.5]{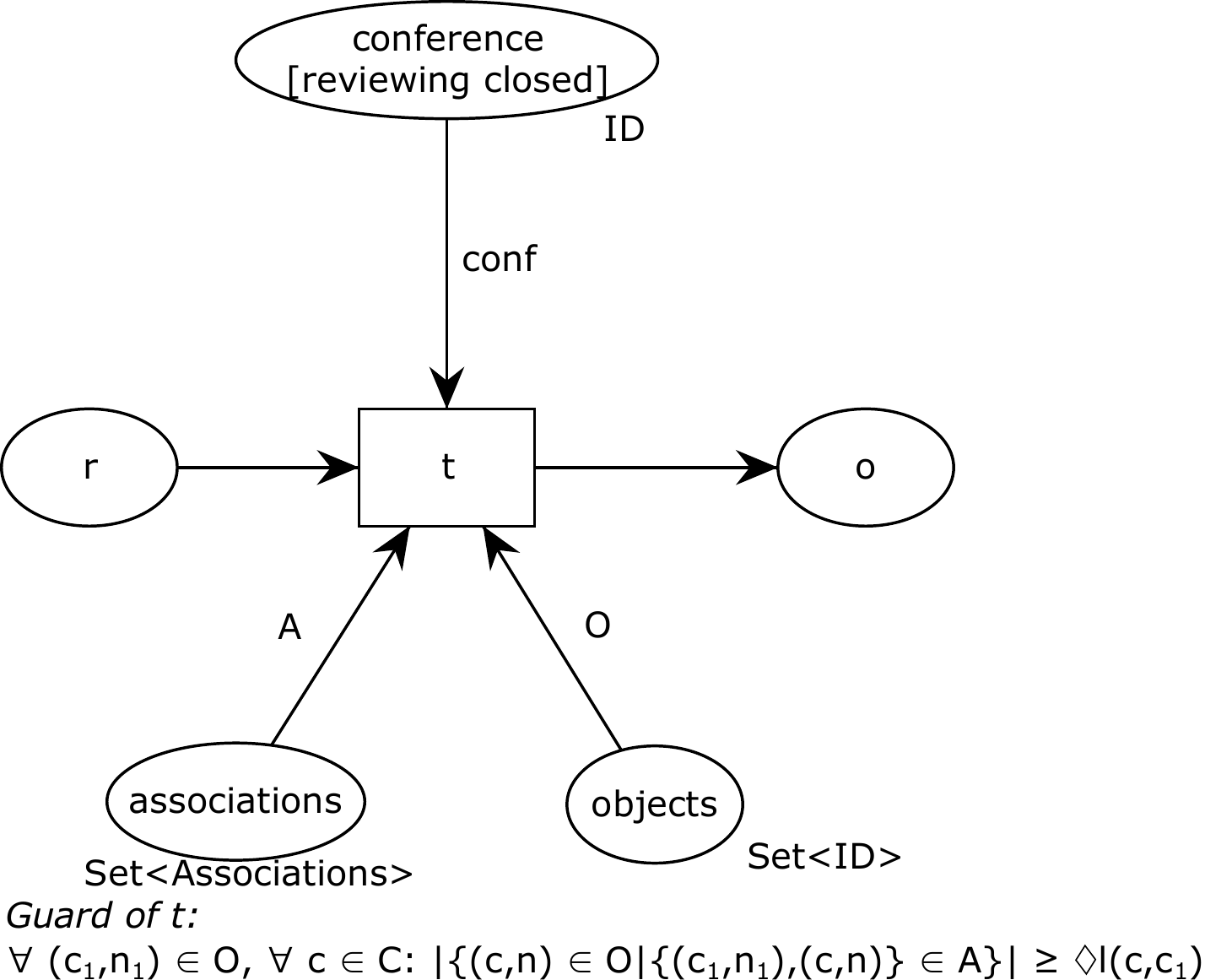}
	\caption{A transition for the termination condition $\{con\!f\!erence[reviewingClosed]\}$. Firing the transition closes the case by moving a token from \emph{r} to \emph{o}. The guard asserts all goal cardinality constraints. $C$ is the set of classes in the domain model (e.g., \emph{Conference, Paper, AuthorTeam, Review}, and \emph{Decision}).}
	\label{fig:t_formal}
\end{figure}
The case can terminate once the conference is in state \textit{reviewing closed} and all goal cardinalities are satisfied (\autoref{fig:t_formal}).
This is represented by one transition that moves a token from \emph{r} (case is running) to \emph{o} (case has been closed).
It furthermore \emph{consumes} the set of objects and associations, as well as a token for the required object configuration \emph{conference}[\emph{reviewing closed}].
Since no further activities should be executed, the tokens are not reproduced.
The guard checks the associations (A) of each object (in O) and verifies whether the goal cardinality constraints are met.

	\section{Implementation and Discussion}
\label{sec:discussion}
The encoding of case models into CPNs detailed in \autoref{sec:execution_semantics} provides a basis for full-fledged enactment of fCM models. In this respect, we present prototypes of i) a compiler translating fCM models to CPNs compatible with CPNTools and of ii) an execution engine.
Prototypes, instructions, and screencast are available at {\small\url{https://github.com/bptlab/fcm2cpn/tree/caise}} and {\small\url{https://github.com/bptlab/fCM-Engine/tree/caise}}.

\begin{figure}[htb]
	\centering
	\includegraphics[]{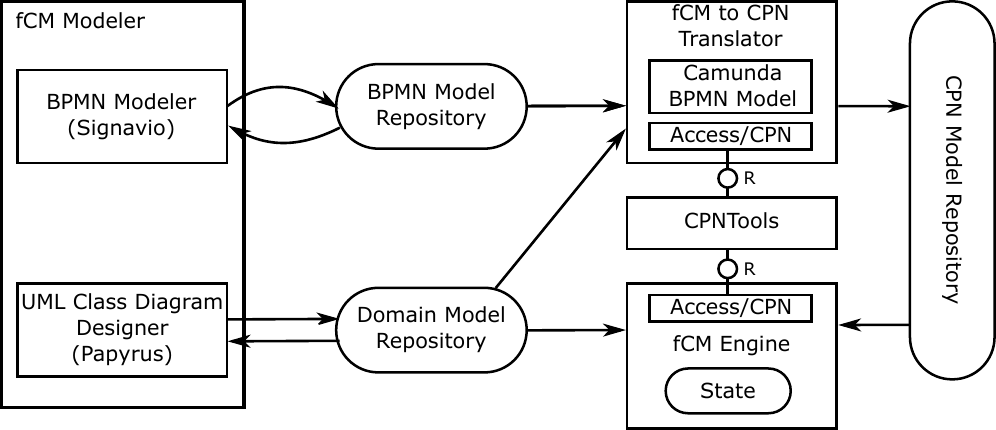}
	\caption{FMC model of the architecture. The compiler reads a case model consisting of a UML class diagram and BPMN file containing fragments and produces a CPN. The engine uses CPNTools to execute cases. Additional information, such as detailed information about classes and attribute information, are stored in the state.}
	\label{fig:architecture}
\end{figure}

Figure~\ref{fig:architecture} shows the architecture of our implementation.
Case models can be created using common tools for BPMN and UML modeling.
Given a BPMN file describing a set of fragments and a domain model (UML Class diagram without hierarchies) describing the classes of objects used in the case model, the compiler creates a CPN file using Access/CPN.
It established a connection between the two inputs by matching the names of data objects in the fragments to the names of classes in the domain model.
We assume that the fragments conform to the OLC,s enabling us to extract the OLCs from the fragments.
\begin{figure}[htb]
\begin{floatrow}
	\capbtabbox{%
		\scriptsize
		\begin{tabular}{r@{\hspace{5mm}}l}
			\toprule
			r & () \\ \midrule
			paper[in\_review] & (Paper, 0) \\ \midrule
			review[available] & (Review, 0) \\
			& (Review, 1) \\ \midrule
			associations & [((Paper, 0), (Review, 0)), \\
			& ((Paper, 0), (Review, 1))] \\ \midrule
			objects & [(Paper, 0), (Review, 0),\\
			& (Review, 1)] \\ \midrule
			cnt review & 1 \\
			\bottomrule
		\end{tabular}
	}{
		\caption{Relevant marking for CPN, see \autoref{fig:fe_formal} for place names.}
		\label{tbl:marking}
	}\hfill
	\ffigbox{
		\includegraphics[width=.5\textwidth]{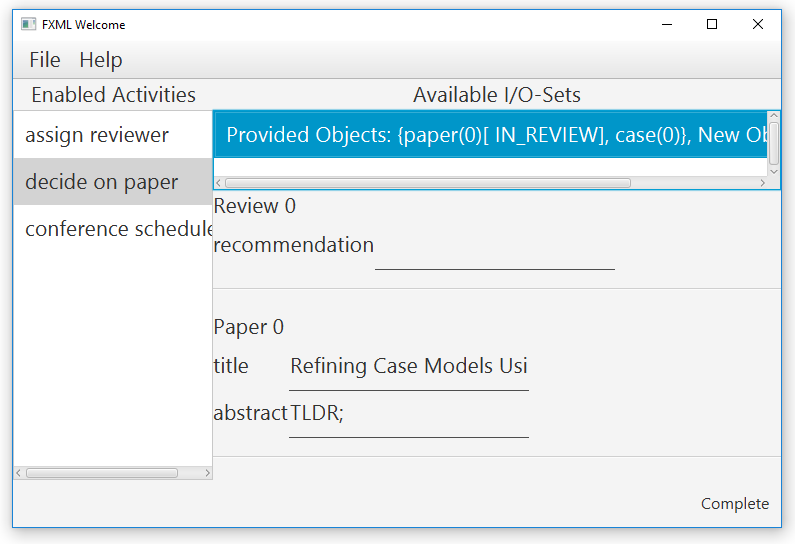}
		\vspace{-1.5em}
	}{
		\caption{The engine depicting the state from \autoref{tbl:marking}. A new review is created.}
		\label{fig:screenshot}
	}
\end{floatrow}
\end{figure}

\begin{wraptable}[15]{r}{.6\textwidth}
	\centering
	\scriptsize
	\caption{Requirements of knowledge-intensive processes~\cite{diCiccio2015} fully (+) or partly ($\circ$) satisfied by fCM ($+$). We do not include requirements regarding user management and adaptability.}
	\label{tab:requirements}
	\begin{tabular}{l@{\hspace{5mm}}l@{\hspace{5mm}}r}
		\toprule
		R1 & Data Modeling & $+$ \\
		\midrule
		R4 & Synchronized Data Access & $\circ$ \\
		\midrule
		R5 & Data-driven Actions & $+$ \\
		\midrule
		R7 & Formalized Rules and Constrains & $\circ$ \\
		\midrule
		R9 & Goal Modeling & $+$ \\
		\midrule
		R11 & Support for Different Modeling Styles & $+$ \\
		\midrule
		R12 & Visibility of the Process Knowledge & $+$ \\
		\midrule
		R13 & Flexible Process Execution & $+$ \\
		\midrule
		R24 & Capture and Model External Events & $+$ \\
		\bottomrule
	\end{tabular}
\end{wraptable}
The fCM engine takes the CPN and the corresponding domain model.
The engine communicates with CPNTools to determine the enabled activity and event instances.
CPNTools has tokens as described in \autoref{sec:execution_semantics}.
However, it does not hold any information about data objects' attributes and their values.
This information is stored separately in the engine.
The domain model is parsed to generate forms for data objects.
When a user selects an enabled activity, the engine displays the available input-output set combinations.
The user can choose one, and the corresponding form is displayed (see \autoref{fig:screenshot}).
Once the user completes an action, the engine updates the objects and instructs CPNTools to execute the respective transition.
Currently, valuations and tokens are volatile; once the engine is closed, all progress will be lost.
We made this decision to keep the engine simple.
However, persistent storage, such as a relational database, could be used to store both the state of the CPN and attributes with their values.

Tools are important to deal with the complexity of the approach and the domain.
Our version of fCM contains additional hidden dependencies~\cite{green1989}:
changing one part of a domain model (e.g., cardinality constraints) may require adaptation of other parts (e.g., the inputs of activities).
Additionally, the execution of fragments depends implicitly on the effects that other fragments induce over data objects and their associations.
Well-designed tools for modeling, verification, and execution can compensate for some challenges posed by the language~\cite{green1989}.
The presented engine guides users and makes it optional to interpret the intricate models manually at run-time.
Furthermore, fCM with its details may prove to be a proper low-code solution used by trained engineers rather than business users to design, implement, and monitor knowledge-intensive processes.
However, this work focuses on the concepts and semantics.
Notation and usability are left for future work.
We added domain models with associations and cardinality constraints to case models satisfying requirements for knowledge-intensive processes (cf.~\cite{diCiccio2015}, \autoref{tab:requirements}):
the process model integrates different perspectives each can be used according to their strengths.
Furthermore, the process is data-driven---available objects and cardinality constraints determine the enabled activities.
Finally, data objects are captured in detail with states, associations, and constraints.

Data cardinality constraints are crucial for formal verification.
In general, verifying models combining processes and data is undecidable~\cite{CaDM13}, in particular, because the process may operate on unboundedly many data objects.
As shown in~\cite{MonC16}, a data-aware version of soundness and other data-aware temporal properties can be verified if cardinality constraints with suitable upper bounds and adequately restructured queries are employed.
This work takes a first step towards applying techniques studied in \cite{MonC16} and related approaches to the verification of fCM models, which is one of the next steps we want to take.
	\section{Related Work}
\label{sec:related_work}
We briefly present related work in three categories: fragment-based case management, alternative modeling approaches, and formal models/execution semantics.

fCM is a production case management approach introduced by Meyer et al.~\cite{meyer2014}.
Hewelt and Weske describe fCM, focusing on the language and syntax~\cite{hewelt2016}.
Consecutive works propose methods for eliciting~\cite{hewelt2020} and verifying~\cite{holfter2019} case models.
These works do not consider associations or cardinality constraints.
We add these to fCM, strengthening the link between data and behavior. 

Data-centric process modeling approaches have a long history.
Recently, Steinau et al. provided an overview of such approaches~\cite{steinau2019}.
An early approach is MERODE, which interprets data models with existential associations as a behavioral specification~\cite{snoeck2014}.
Guard stage milestone~\cite{hull2010} and CMMN~\cite{CMMN} arrange activities in stages with data pre- and post-conditions.
PHILharmonicFlows~\cite{kunzle2011} models OLCs, which can be orchestrated into larger processes.
Associations and cardinalities can be used to synchronize transitions between sets of objects~\cite{steinau2018}
In contrast to PHILharmonicFlows and MERODE, fCM focuses on case management.
It is a hybrid approach that separates domain and process modeling.

Rather than specifying process variants imperatively, declarative models define constraints to capture multi-variant processes concisely.
Two prominent declarative languages are DECLARE~\cite{pesci2007} and DCR-Graphs~\cite{hildebrandt2010}, both use variants of linear temporal logics.
While declarative modeling is powerful to model constraints from laws, guidelines, and other rules, imperative models are preferred when explicit control and data flow are important.
DCR-KiPM~\cite{santoro2019} is a hybrid approach combining declarative DCR-Graphs with an imperative notion called KiPM to model knowledge-intensive processes.
Object-centric behavioral constraints combine declarative process models with data models~\cite{aalst2017}.
Activities and LTL rules over finite traces are quantified based on data objects and their associations
Cardinality constraints exist in two flavors: they have to hold globally or eventually.
However, no case exists.
fCM achieves flexibility by combining fragments, modeled using elements within the BPMN tradition, dynamically.
In fCM, declarative rules can be asserted through compliance checking~\cite{holfter2019}.

Formal execution semantics are broadly applied in BPM.
Dijkman et al. present a Petri net formalization for BPMN models enabling analysis and verification~\cite{dijkman2008}.
Proclets~\cite{aalst2000} are an extension of Petri nets: a process is composed of multiple workflow modules that contain synchronization points.
They allow modeling various kinds of processes, such as choreographies and data-centric processes.
Fahland introduces cardinality constraints for the synchronization points enabling many-to-many interactions~\cite{fahland2019}.
DB-Nets assure that a process adheres to constraints given by a database (such as primary and foreign key constraints), by incorporating transactional semantics~\cite{montali2017}.
Catalog nets combine Petri nets with queryable read-only databases to support synchronization among cases~\cite{ghilardi2020}.
Translating fCM to these formalisms would be interesting to connect the two trends of research.
Object-centric Petri nets have been used to describe processes mined from object-centric event logs\cite{aalst2020}.
It would be interesting to see mining algorithms targeting fCM.
Our semantics are defined using Petri nets, but users only interact with the fCM model, which hides the formalism's complexity.
	\section{Conclusion}
\label{sec:conclusion}
Knowledge-intensive processes demand flexibility and data-driven actions.
Also, modeling approaches should incorporate various perspectives through suited  languages~\cite{diCiccio2015}.
In this paper, we integrated activity-centric process fragments with domain models comprising associations and cardinality constraints.
While we use fCM models, our work can be applied to traditional processes (e.g., modeled using BPMN) to determine bounds for loops and multi-instance activities.

We showed that object cardinalities influence the behavior of processes.
Activities are enabled depending on whether certain cardinality constraints hold.
For example, a paper may only be rejected or accepted if three reviews exist (requirement).
On the other hand, only one decision is made for each paper (bound).
Furthermore, cardinality constraints may require batch processing where multiple objects of the same type are accessed simultaneously.
Additionally, cardinality constraints can refine the goal definition of a case.
Not only do objects need to progress into specific states, but they also need to exist in specified quantities.
This allows defining knowledge-intensive processes more comprehensively.

We provide a formal semantics using CPNs respecting the behavioral implications of cardinalities.
The semantics is the underpinning for many BPM related tasks, such as process execution, verification, modeling, and mining.
Due to the intricate nature of knowledge-intensive processes, tools are important.
During modeling, a tool may highlight hidden dependencies and aid with the verification.
At run-time, engines may guide the knowledge workers making it unnecessary to interpret the model manually.
Alternatively, process monitoring may check compliant actions.
While we only focus on the definition of the formal semantics, we lay an important foundation for these future tasks.

We do not consider inheritance.
Class hierarchies where associations and cardinality constraints are the same on all levels do not require any changes.
However, specializing and generalizing associations including cardinality constraints and advanced modeling concepts, e.g., the phase pattern, are interesting for future work and require investigation of conceptual models and algorithms.

Future work may also investigate the role of knowledge workers in detail:
they may require relaxed constraints or the ability to adapt the process ad hoc.

\subsubsection{Acknowledgments.} We thank Leon Bein for his work on the prototypes.
	\bibliographystyle{splncs04}
	\bibliography{bib/bibfile}
\end{document}